\documentclass{vgtc}
\DeclareUnicodeCharacter{202F}{\,}
\usepackage{amsmath}
\usepackage{graphicx}
\usepackage{stfloats} %
\usepackage{caption}
\usepackage{textalpha} 

\usepackage{enumitem}
\usepackage{makecell}
\usepackage{array}

\graphicspath{{figures/}{pictures/}{images/}{./}} 

\usepackage{times}                     

\usepackage{tabu}                      
\usepackage{booktabs}                  
\usepackage{lipsum}                    
\usepackage{mwe}                       

\usepackage{mathptmx}                  

\onlineid{2057}

\vgtccategory{Research}

\vgtcinsertpkg

\title{Merging Bodies, Dividing Conflict: Body-Swapping in Mixed Reality Increases Closeness Yet Weakens the Joint Simon Effect}

\author{Yuan He\thanks{e-mail: hey3@tcd.ie}\\ %
        \scriptsize Trinity College Dublin %
\and Brendan Rooney\thanks{e-mail: brendan.rooney@ucd.ie}\\ %
     \scriptsize University College Dublin %
\and Rachel McDonnell\thanks{e-mail: RAMCDONN@tcd.ie}\\ %
     \parbox{1.4in}{\scriptsize \centering Trinity College Dublin}}

\teaser{
  \centering
  \includegraphics[width=1\linewidth]{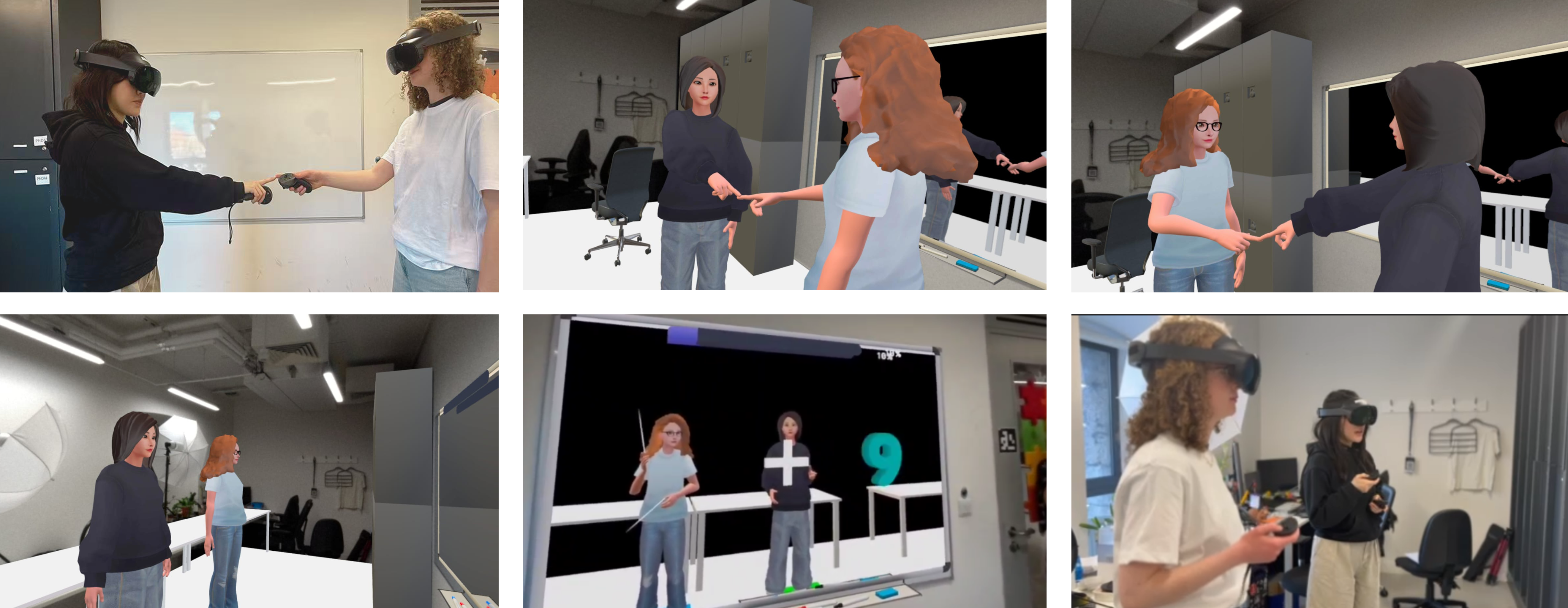}
  \caption{Top row, left to right: (1) Real-world view of the participant (white T-shirt) and experimenter (black hoodie) touching index fingers; (2) the corresponding Mixed Reality (MR) view with each person’s own avatar; and (3) the MR scene after performing a body swap. Bottom row, left to right: (4) the participant and experimenter standing in their positions for the Joint Simon Task under the swapped condition; (5) an in-task MR screenshot of the Joint Simon Task; and (6) the real-world view of both individuals performing the Simon Task.}
  \label{fig:teaser}
}

\abstract{
Mixed Reality (MR) presents novel opportunities to investigate how individuals perceive themselves and others during shared, augmented experiences within a common physical environment. Previous research has demonstrated that users can embody avatars in MR, temporarily extending their sense of self. However, there has been limited exploration of body-swapping, a condition in which two individuals simultaneously inhabit each other’s avatars, and its potential effects on social interaction in immersive environments. To address this gap, we adapted the Joint Simon Task (JST), a well-established implicit paradigm, to examine how body-swapping influences the cognitive and perceptual boundaries between self and other. Our results indicate that body-swapping led participants to experience themselves and their partner as functioning like a single, unified system, as in two bodies operating as one agent. This suggests possible cognitive and perceptual changes that go beyond simple collaboration. Our findings have significant implications for the design of MR systems intended to support collaboration, empathy, social learning, and therapeutic interventions through shared embodiment.
    
} 

\keywords{Mixed Reality, Body Swap, Self-concept Clarity, Inclusion of Other in the Self, Joint Simon Task, Social-presence, Embodiment.}

\begin{document}

\maketitle

\section{Introduction}
Mixed Reality (MR) systems increasingly allow multiple users to perceive digital overlays while remaining aware of the surrounding physical space. When those augmentations include self‑avatars, users can go beyond simple co-presence and experience variations of bodily representations that are difficult or impossible in the real world. Classic laboratory works show that synchronized visual and sensory feedback can shift body ownership to a different form, sometimes of a different age, race, or even species, within seconds \cite{petkova2008if}. Yet most studies stop at confirming the illusion’s validity; only a few examine how altered embodiment affects real‑time coordination among co‑located users.

A valuable lens on such coordination is the Simon task. In the standard Simon task, participants respond to non-spatial features of stimuli (e.g., shape) using spatially defined responses (e.g., left button for cube, right for pyramid). Although stimulus location is irrelevant, people typically respond more slowly when the stimulus side conflicts with the responding hand (e.g. cube appears on the right), a phenomenon known as the Simon Effect \cite{simon1967auditory}. The effect usually disappears if a participant performs a Go/No-Go version of the task, that is, they only respond to one stimulus (e.g., press the button when the cube appears and do not press when the pyramid shows up). However, it reemerges when two people share the task to respond to only one feature each. In this joint task, the Joint Simon Effect (JSE) still emerges, where responses are quicker for stimuli that appear on the same side as the responding participant. The spatial interference associated with another person’s response suggests that they have unwittingly encoded not only their own rule but also their partner’s, treating the dyad like a single bimanual actor \cite{dolk2014joint}.  The magnitude of the Joint Simon Effect (JSE) grows with perceived social closeness, implying that it can serve as an implicit indicator of interpersonal closeness \cite{shafaei2020effect}.

We propose that reciprocal body-swapping lets two users view their own virtual bodies from a third-person perspective while simultaneously controlling each other’s avatars from a first-person view. This manipulation may prompt the brain to incorporate the partner’s body into the sense of self, softening the boundary between self and other. Such a fusion of body representations could reorganize the cognitive mechanisms that give rise to the Joint Simon Effect (JSE). Central to these mechanisms is action co-representation: each participant automatically builds an internal model of the partner’s forthcoming movements and keeps it in parallel with their own action plan. Although recent prototypes allow two users to exchange avatars in virtual environments \cite{dollinger2024virtual}, no study to date has examined whether reciprocal body-swapping amplifies this shared coding of action during real-time interaction.

To investigate this question, we developed a four-condition Simon task protocol. Each participant completed a standard Simon task, a Go/No-Go task, as well as a Joint Simon Task under both non-swapped and reciprocal body-swapping conditions. Including the two baseline tasks allowed us to control for general effects such as sensorimotor adaptation or fatigue, ensuring that any changes in the Joint Simon Effect could be attributed specifically to the body-swapping manipulation.

We addressed four intertwined research questions:
\begin{itemize}
    \item \textbf{RQ1}: Does reciprocal body‑swapping change the Joint Simon Effect (JSE), indicating that partners act as a unified system?

    \item \textbf{RQ2}: Given prior evidence that JSE magnitude reflects interpersonal closeness, does swapping influence perceived closeness and, in turn, the JSE?
    
    \item \textbf{RQ3}: Does reciprocal body‑swapping alter specific components of sense of embodiment (body‑ownership, agency, self‑location), and do those alterations predict the magnitude of the JSE?

    \item \textbf{RQ4}: Do stable individual traits, such as clarity of own self-concept and personality, moderate the extent to which body-swapping influences the JSE?
\end{itemize}

This study makes the following contributions. Methodologically, we introduce the first MR‑based Joint Simon Task with real‑time reciprocal avatar swapping, allowing for fine-grained assessment of action co-representation under dynamic embodiment conditions. Empirically, we show that body‑swapping systematically attenuates the JSE, evidencing a reorganization of self-other coding. Theoretically, we demonstrate that the magnitude of ownership loss, not interpersonal closeness, predicts the behavioral attenuation, refining accounts of referential coding. Practically, our results offer guidance for the design of MR systems aimed at enhancing collaboration, perspective-taking, and therapeutic engagement by modulating the perceived boundary between self and other.

\section{Related Work}
\subsection{Virtual Body-Swapping}
Virtual body-swapping refers to using immersive technology to induce the illusion of owning a different body, typically by synchronizing visual perspective and touch to swap one’s perceived body with another’s \cite{petkova2008if}. In a seminal study, Petkova and Ehrsson used head-mounted cameras that allowed each participant to look down and see the other’s torso; synchronous stroking generated strong ownership of the partner’s body along with a shift in self-location. Later work showed that the illusion can also be triggered with fully virtual avatars \cite{osimo2015conversations}, and that embodying a child-sized or different-race body can alter perception and implicit attitudes \cite{banakou2013illusory,banakou2018virtually,peck2013putting}.

Most studies have focused on one-way swaps, where users embody prerecorded or computer-controlled bodies and complete offline tasks or questionnaires. Only a few systems support reciprocal, real-time swapping, in which two users control each other’s avatars simultaneously. Examples include BeAnotherLab’s “Machine to Be Another” \cite{ventura2024being} and the VR body-swapping system by Dollinger et al. \cite{dollinger2024virtual}.

In mixed reality (MR), body-swapping remains largely unexplored. Genay et al. found that overlaying live video of a partner’s limbs onto the user’s field of view could evoke strong ownership, though the social consequences were not assessed. No prior work has tested whether reciprocal body-swapping in MR alters implicit action coupling, such as the joint Simon effect that reflects action co-representation. Addressing this gap is critical to understanding how shared embodiment might support collaboration, empathy, and rehabilitation in everyday settings. The present study integrates a real-time MR body-swap with a joint Simon task to examine how swapping two full avatars modulates self–other coding during live interaction.

\subsection{Joint Action and the Joint Simon Task}

During joint action, collaborators form action co-representations by running an internal model of a partner’s upcoming movement alongside their own motor plan \cite{atmaca2008action}. A widely used behavioral index of this shared coding is the Joint Simon Effect (JSE).

In the standard Simon task, participants classify a non-spatial feature, such as shape or color, using a spatially defined response, typically pressing a left or right key. Responses are slower when the stimulus appears on the side opposite to the correct response. This delay is known as the Simon Effect \cite{simon1967auditory}.The effect disappears in a Go/No-Go version of the task, where a single participant responds to only one stimulus type and ignores the other. In this version, spatial location is no longer relevant, and thus the conflict does not arise. However, when this Go/No-Go structure is shared between two participants, with each person responding to only one stimulus type, the effect returns. This version is known as the Joint Simon Task. The reappearance of the effect suggests that each individual not only applies their own response rule but also represents the partner’s rule, as if forming a shared action system \cite{sebanz2003representing}.

Two theories explain the JSE. The task co-representation account suggests that partners store both stimulus–response mappings as a bimanual set. The referential coding account argues that the cognitive system uses spatial location to distinguish self-generated actions from external ones \cite{dolk2014joint}. Both accounts predict a reduced JSE when the boundary between self and other actions becomes unclear.

Laboratory evidence supports this view. For instance, when the rubber-hand illusion induces ownership over a partner’s hand, Simon interference is greatly reduced or disappears entirely \cite{dolk2011social}. Reciprocal body-swapping may push this merger further, as participants exchange full body representations. Once the swap becomes convincing, the partner’s keypress is perceived as originating from the same body and no longer serves as an external cue, thereby diminishing the JSE.

Based on this rationale, we formulated the following hypothesis, aligned with RQ1:

\textbf{H1.} Reciprocal body swapping in mixed reality will attenuate the Joint Simon Effect compared to the Self-Avatar condition, indicating that avatar swapping restructures action co-representation.

\subsection{Social Presence and Interpersonal Closeness in Joint Action}

The magnitude of the Joint Simon Effect (JSE) reflects how strongly individuals integrate a partner’s stimulus–response rule into their own \cite{ford2015exploring}. Prior research shows that this effect depends on both the identity of the partner and their perceived presence \cite{colzato2012up, mcclung2013group}. Friendly or in-group partners yield larger JSEs than unfriendly or out-group partners \cite{muller2011perspective}. This pattern extends to real friendships: Ford and Aberdein found that friend dyads exhibited greater JSE than strangers, with the effect scaling with individual empathy scores \cite{ford2015exploring}.

Shafaei et al. also reported a positive correlation between Joint Simon Effect (JSE) magnitude and \textbf{Inclusion of Other in the Self (IOS)} scores in both adolescents and adults, confirming that perceived closeness reliably predicts action co-representation \cite{shafaei2020effect}. Activities that promote shared identity or interdependence further amplify the effect \cite{wahn2021humans}, whereas describing the co-actor as a mechanical device eliminates it \cite{stenzel2012humanoid}. When agency cues are introduced, such as by labeling the device as autonomous and intelligent, the Simon effect re-emerges \cite{pfister2014action}. These findings suggest that social presence, defined as the perception of another intentional agent being present, and interpersonal closeness both contribute to the modulation of action co-representation.

Immersive technologies reinforce this pattern. When physically distant partners share a virtual space via avatars, the JSE reappears despite a lack of co-location \cite{harada2025effect}. Avatar fidelity matters: full-body humanoid avatars elicit strong JSEs, while disembodied hands produce weaker effects \cite{li2024social}. Minimal cues may suffice if they convincingly signal agency. Spatialized footsteps from the partner’s location restore the JSE, whereas non-spatial sounds do not \cite{kiridoshi2022spatial}. In contrast, thin media like video calls fail to elicit the JSE despite enhancing general social engagement \cite{sobel2024joint}, highlighting the importance of presence and spatial alignment for action co-representation.

Mixed reality provides a coherent spatial frame and rich social cues, making it well suited to test how reciprocal body-swapping reshapes the link between explicit social experience and implicit joint action. Prior studies predict a positive relationship between closeness and JSE magnitude. However, as discussed in Section 2.2, body-swapping might attenuate the JSE. To examine whether this link persists under both conditions, we propose:

\textbf{H2a.} In the Self-Avatar condition, higher interpersonal closeness (IOS) will predict larger JSE magnitudes.

\textbf{H2b.} In the Swapped-Avatar condition, higher interpersonal closeness (IOS) will predict larger JSE magnitudes.

In addition, we test whether changes in social‑presence subscales (Co‑presence, Attentional Allocation, Behavioral Interdependence) also predict the JSE; no directional hypothesis is specified given mixed prior findings.

\subsection{Sense of Embodiment (SoE) in Mixed Reality}

\textbf{The sense of embodiment (SoE)} includes three partially dissociable components: body ownership, agency, and self-location \cite{kilteni2012sense}. Experiments show that each can be selectively disrupted. For example, passive finger displacement removes agency but preserves ownership \cite{kalckert2012moving}, while third-person full-body perspectives shift self-location without transferring ownership to the distant avatar \cite{aspell2012multisensory}. These findings suggest that SoE arises from multiple multisensory processes rather than a single unified mechanism.

Although early embodiment research focused on fully immersive virtual reality, mixed reality (MR) can produce illusions of equal or greater strength when visual and proprioceptive cues remain spatially aligned \cite{genay2021being}. Several studies report no significant differences in ownership ratings across virtual, augmented, and physical settings \cite{vskola2016examining,wolf2020body}. Genay et al. further found stronger ownership for virtual hands in combined real–virtual environments than in purely virtual ones \cite{genay2021virtual}. These results indicate that viewing the physical environment can enhance embodiment illusions, especially when sensorimotor congruency is preserved.

MR also strengthens the plausibility illusion, the sense that virtual events are unfolding here and now \cite{slater2009place,slater2022separate}. When virtual actions follow real-world physics and interact coherently with the surroundings, users experience greater plausibility \cite{westermeier2023exploring}. Compared to enclosed VR systems, MR supports this coherence and reduces cybersickness \cite{kirollos2023comparing}. It therefore serves as a suitable testbed for investigating the sense of embodiment.

We hypothesize that changes in each SoE sub-component will correspond to changes in the Joint Simon Effect (JSE) following reciprocal body-swapping. Specifically:

\textbf{H3a (Body ownership):} A larger decrease in body‑ownership after swapping will predict a larger reduction in JSE.

\textbf{H3b (Agency):} Agency will remain unaffected by body-swapping, as the level of control does not change.

\textbf{H3c (Self-location):} A greater out‑of‑body shift (increase in self‑location score) will predict a larger reduction in JSE.

Together, H3a–H3c test whether subjective embodiment changes explain the recoding of action representations indexed by the Joint Simon Task.

\subsection{Self-Concept Clarity and Personality Traits Modulate Sense of Embodiment}

Individual differences in self-identity and personality significantly influence how users experience embodiment and collaborate in immersive environments. For example, \textbf{self-concept clarity (SCC)}, defined as the extent to which one’s self-concept is internally consistent and clearly structured, has been shown to affect susceptibility to embodiment illusions \cite{campbell1996self}. Individuals with lower SCC are more likely to experience “body-transfer” illusions, more easily feeling ownership over artificial or virtual bodies. In contrast, those with higher SCC tend to maintain a more stable bodily self and exhibit weaker illusion effects \cite{krol2020self}.

\textbf{Personality traits} also shape embodiment and social presence. Extraversion and openness to experience have been linked to greater presence and comfort in virtual contexts \cite{dewez2019influence, katifori2022exploring}, which may enhance collaboration. Empathy-related traits, such as perspective-taking, predict stronger presence and better task performance \cite{katifori2022exploring}. Conversely, neuroticism may increase anxiety in immersive scenarios, reducing presence and hindering coordination. Agreeableness has more complex effects. While it typically fosters trust and cooperation, excessive agreeableness may promote groupthink. Notably, dyads composed of two low-agreeableness individuals achieved the greatest performance gains in a virtual task, likely due to more direct and critical communication \cite{zhu2024benefits}.

These findings suggest that the Swapped-Avatar condition may influence individuals differently depending on their trait profiles. Participants with low SCC or high openness may find it easier to adopt the new avatar, leading to greater changes in perceived self–other boundaries and stronger modulation of the Joint Simon Effect (JSE). In contrast, individuals with high SCC or high neuroticism may resist the illusion, showing reduced behavioral change.

Based on this rationale, we formulate the following hypothesis.

\textbf{H4.} Higher self-concept clarity will predict a greater swap-induced increase in interpersonal closeness ($\Delta$IOS). No a priori predictions are made regarding embodiment measures or the Joint Simon Effect.

We will also explore whether Big Five traits, including Extraversion, Agreeableness, Conscientiousness, Neuroticism, and Openness to Experience, are associated with $\Delta$IOS or moderate the impact of swapping on the JSE. These analyses are exploratory and do not include directional hypotheses.

Together, these hypotheses connect stable individual traits to both the social and behavioral outcomes of avatar swapping, thereby addressing \textbf{RQ4}.

\section{Methods}

\subsection{Participants}

An a priori power analysis (G*Power 3.1, repeated-measures ANOVA, within-subject factors, effect size $f = 0.25$, $\alpha = 0.05$, $1 - \beta = 0.80$, assumed correlation = 0.5) indicated that a sample of 24 participants was required. We recruited 33 adults (21 male, 11 female, 1 non-binary; $M = 29.7$ years, $SD = 6.9$), all of whom reported normal or corrected-to-normal vision. Each participant was paired with the experimenter to complete the study.

Written informed consent was obtained prior to participation, and the study was approved by the Trinity College Dublin Research Ethics Committee. All participants received a 10-euro book voucher as compensation.

All sessions took place in a 2.5 m × 5 m laboratory space cleared of obstacles. Participants wore a Meta Quest Pro headset (90 Hz, 1800 × 1920 pixels per eye) in color passthrough mode. Virtual content was anchored using the Meta Mixed Reality Utility Kit (MRUK) to ensure geometric stability during movement.

A virtual mirror (1.8 m × 1.2 m) was anchored to the laboratory whiteboard using MRUK and positioned 1.65 m in front of the participant. It displayed a real-time third-person view of the avatar and served as the stimulus panel. A fixation cross and target stimuli were rendered at the center, replicating the layout used in standard Simon tasks. This setup allowed participants to view their embodied form without shifting gaze from the task display.

To minimize occasional mismatches between the physical body and the partner-controlled avatar due to network latency, we recreated key pieces of furniture as 3D models and overlaid them at startup using MRUK. Through the headset’s Physical Environment mapping, we scanned or manually aligned the laboratory desk, floor surface, and a ceiling-high cabinet. These meshes were anchored so their virtual counterparts closely matched the physical furniture. Positioned along typical lines of sight from the torso to the lower limbs, these occluders masked brief exposures of the real body when rapid movement exceeded avatar update speed. To preserve natural lighting and spatial cues, all other elements such as walls, doors, windows, ceiling lights, and objects on the desk remained in passthrough.

Full-body and facial motion were captured using the Meta Movement SDK. Inverse kinematics was computed from head and controller poses, and each avatar mesh was automatically scaled to match the user’s body dimensions. At the beginning of each block, the participant stood on a floor marker, and the avatar root was adjusted to align virtual and real feet, minimizing visuo-proprioceptive conflict.

Multiplayer synchronization used Unity 2022.3 LTS Netcode for GameObjects in host-client mode without an external server. Avatar root and hand poses were streamed at 60 Hz over a dedicated Wi-Fi 6 router, with round-trip latency below 8 milliseconds on the local network.

\begin{figure}[h]
    \centering
    \includegraphics[scale=0.17]{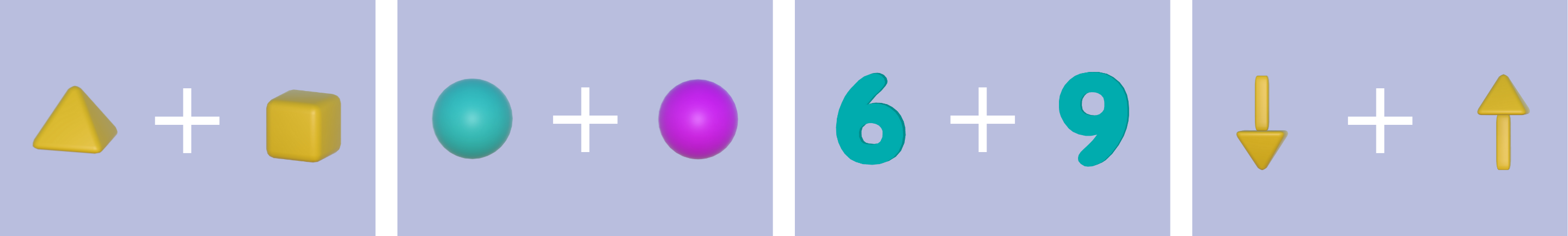}
    \caption{Examples of the four stimulus types: geometric solids, colored spheres, digits, and arrows (left to right).}
    \label{fig:stimuli}
\end{figure}

To introduce variety and prevent over-familiarity, we used four stimulus types: geometric solids, colored spheres, digits, and arrows. As shown in Figure~\ref{fig:stimuli}, each trial began with a 500 ms fixation cross, followed by a 1000 ms stimulus displayed on either side of fixation. Participants had up to 1000 ms to respond using the controller grip button. Responses submitted after this window were classified as misses. A 500 ms blank screen separated each trial to prevent carryover effects.

\begin{figure}[h]
    \centering
    \includegraphics[scale=0.18]{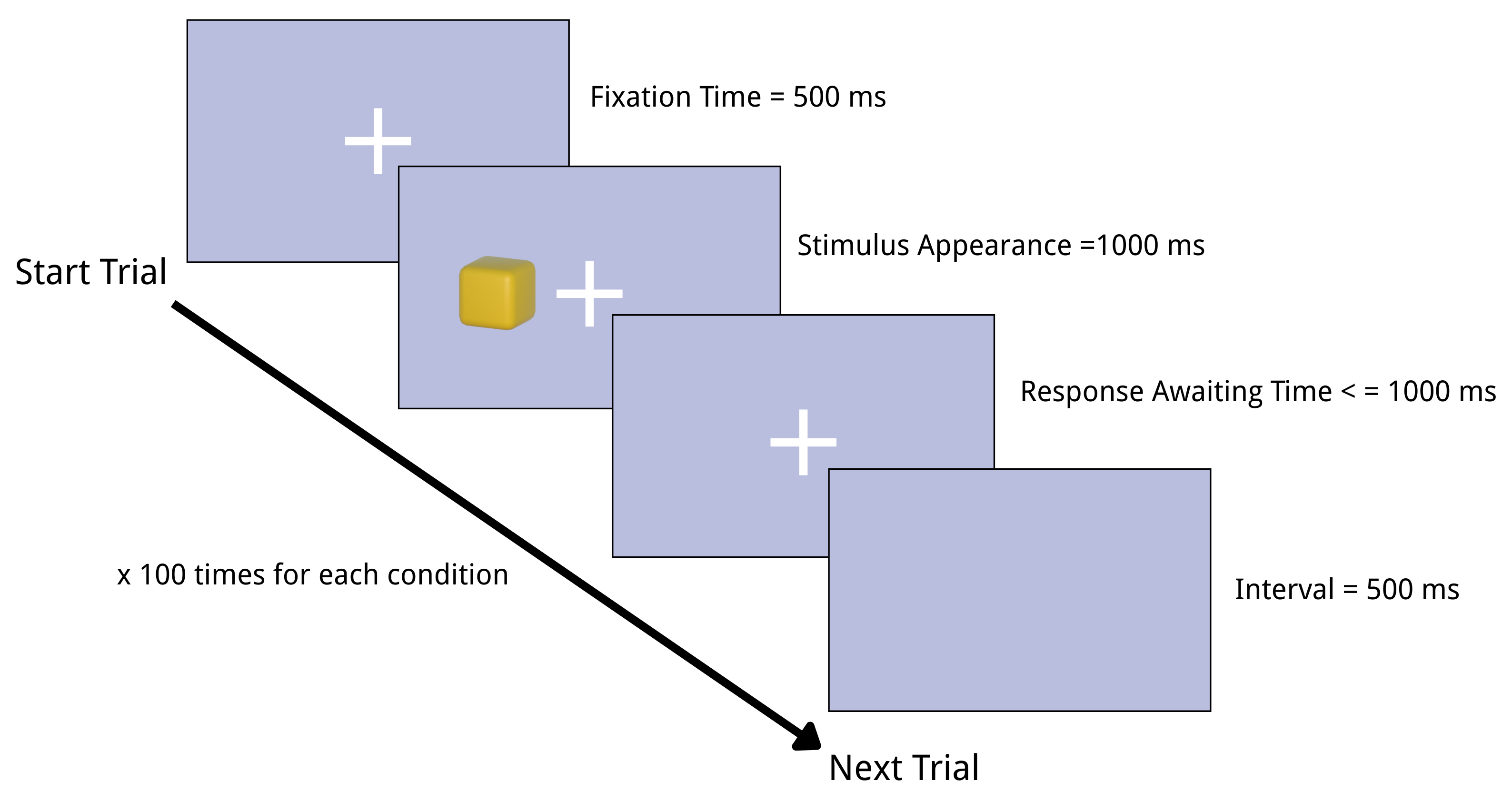}
    \caption{Timeline of a single trial. A 500\,ms fixation cross is followed by a 1000\,ms target stimulus. Participants have up to 1000\,ms to respond, followed by a 500\,ms inter-trial interval.}
    \label{fig:timeline}
\end{figure}

Figure~\ref{fig:timeline} shows the trial sequence. Each experimental condition included 100 trials, with stimulus type and location randomized to minimize order effects. This ensured balanced exposure to compatible and incompatible trials and prevented any single stimulus or side from dominating participant responses.

\subsection{Measures}
The primary outcome measure was performance on the implicit Joint Simon Task. We also collected self-report data to assess participants’ subjective impressions under the two main experimental conditions (Self-Avatar and Swapped-Avatar). Finally, we administered trait-level questionnaires to explore whether individual differences could explain variability in the results.

\subsubsection{Implicit}

The Joint Simon Task served as our implicit behavioral measure. The classic Simon effect refers to faster responses to spatially compatible than incompatible stimuli. This effect typically disappears when a single participant performs a Go/No-Go version. When it reemerges in a shared task, where two individuals split response subsets, the resulting Joint Simon Effect (JSE) reflects co-representation of the partner’s response rules. We quantified the JSE as the difference in mean reaction time between compatible and incompatible trials under each joint condition.

\subsubsection{Experiment Questionnaires}

\begin{figure}[h]
    \centering
    \includegraphics[scale=0.17]{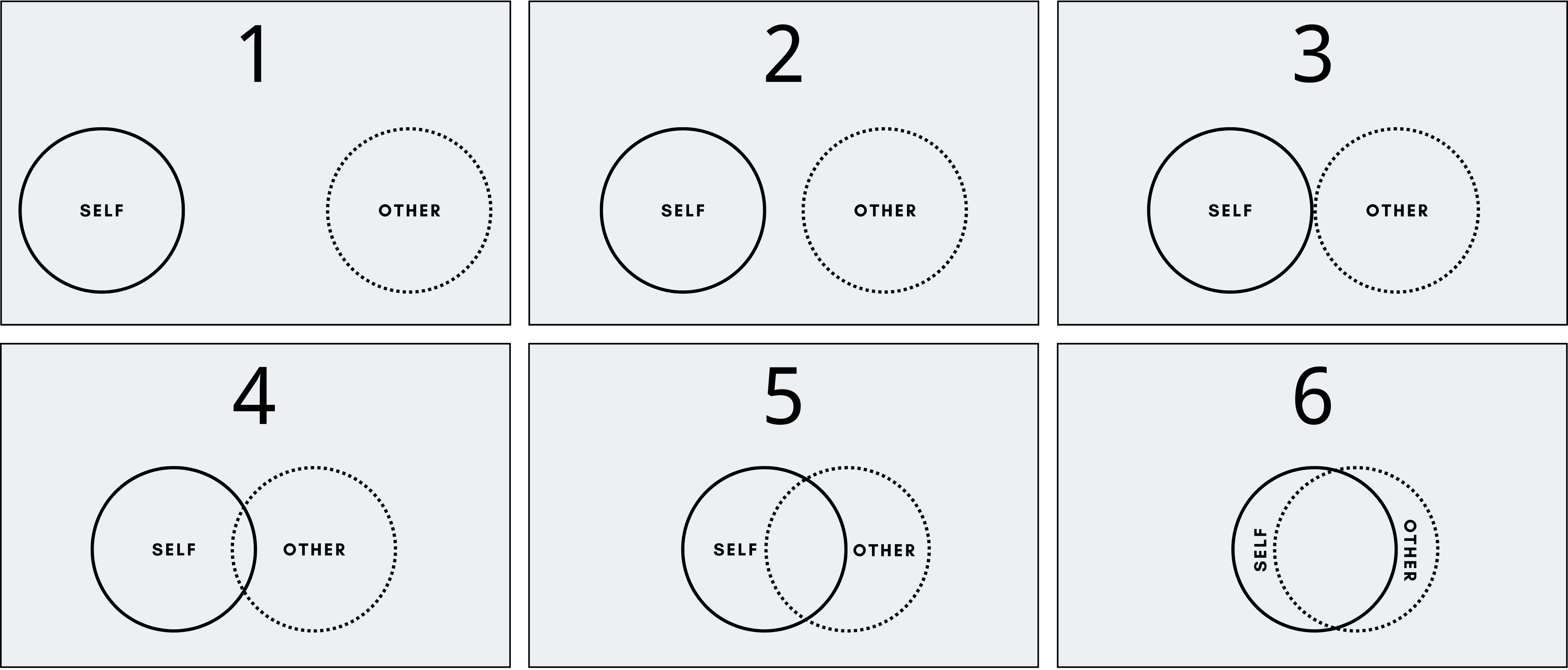}
    \caption{Illustration of the 6-step IOS scale used to measure perceived self–other overlap, ranging from mild separation to deep closeness.}
    \label{fig:ios}
\end{figure}

We used a 6-step version of the Inclusion of Other in the Self (IOS) scale \cite{shafaei2020effect}, adapted from the original 7-step format by Aron et al. \cite{aron1992inclusion}. This version accommodates both perceived closeness and distance. As illustrated in Figure~\ref{fig:ios}, the first one or two circle pairs indicate mild or moderate distance, while higher overlaps represent moderate to extreme closeness. The adjustment captures a wider range of interpersonal perceptions, including potential ambivalence or conflict.

An important component of body-swapping is the \textbf{Sense of Embodiment}, which comprises three subcomponents based on the framework by Kilteni et al. \cite{kilteni2012sense}: \textbf{Ownership}, the feeling that the avatar’s body is one’s own; \textbf{Agency}, the perceived control over the avatar’s movements; and \textbf{Location}, the sense of being spatially situated within the avatar. We adapted specific items for each subscale from the questionnaire by Gonzalez-Franco et al. \cite{gonzalez2018avatar}.

Each subscale included brief statements (e.g., "I felt as if the avatar's body was my own") rated on a Likert scale.

To assess \textbf{Social Presence}, we selected subscales from the Networked Minds questionnaire \cite{biocca2003guide}, including \textbf{Co-presence}, or the awareness of sharing space with another person; \textbf{Attentional Allocation}, the degree of attention directed toward the partner; and \textbf{Perceived Behavioral Interdependence}, the extent to which each person feels that their actions influence and are influenced by the other.

\subsubsection{Trait Questionnaires}

To assess how consistently and confidently participants define themselves, we administered the 12-item \textbf{Self-Concept Clarity (SCC)} scale \cite{campbell1996self}. Participants rated statements such as “My beliefs about myself often conflict with one another” on a 5-point Likert scale, with higher scores indicating greater clarity.

The \textbf{Ten-Item Personality Inventory (TIPI)} \cite{gosling2003ten} provided a brief assessment of the Big Five traits: Extraversion, Agreeableness, Conscientiousness, Neuroticism, and Openness. Each trait was measured using two items, enabling us to examine how personality dimensions might moderate responses to body swapping.

\section{Procedure}

\begin{figure}[h]
    \centering
    \includegraphics[scale=0.2]{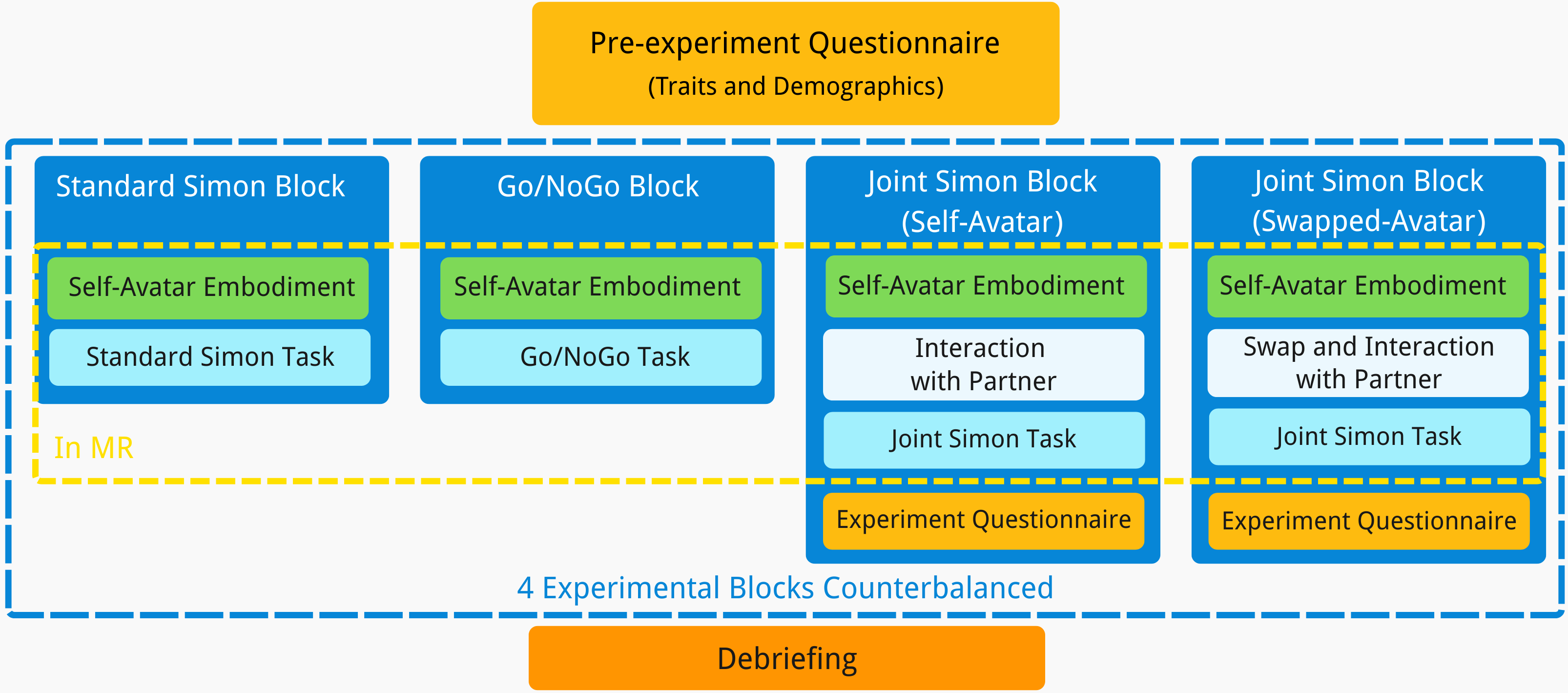}
    \caption{Overview of the experimental procedure.}
    \label{fig:procedure}
    \vspace{-4mm}
\end{figure}

The full procedure of the experiment is illustrated in Figure~\ref{fig:procedure}.

\subsection{Pre-experiment}

Before the experiment, participants used the \textit{readyplayer.me} platform to create a personalized 3D avatar from a frontal photograph. They were asked to choose a white short-sleeve T-shirt, matching the real shirt provided during the session, to minimize discrepancies between virtual and physical appearance. Once submitted, the experimenter completed rigging and integrated the avatar into the experimental platform.

Upon arrival, participants provided written informed consent and completed a brief questionnaire including demographic items, the Self-Concept Clarity (SCC) scale, and the Ten-Item Personality Inventory (TIPI). They then changed into the white T-shirt and wore the \textit{Meta Quest Pro} headset. The experimenter assisted with calibration to ensure accurate display and interaction within the mixed reality environment.

\subsection{Experiment Blocks}
We implemented four experimental conditions, presented in counterbalanced order to reduce learning effects. Two single-participant baselines isolated general factors underlying spatial compatibility: the \textbf{Standard Simon Task}, in which separate left and right hand responses typically produce a robust compatibility effect, and the \textbf{Go/No-Go Task}, in which a single response key largely eliminates it.  

\begin{figure*}[!t]  
  \centering
  \includegraphics[width=\textwidth]{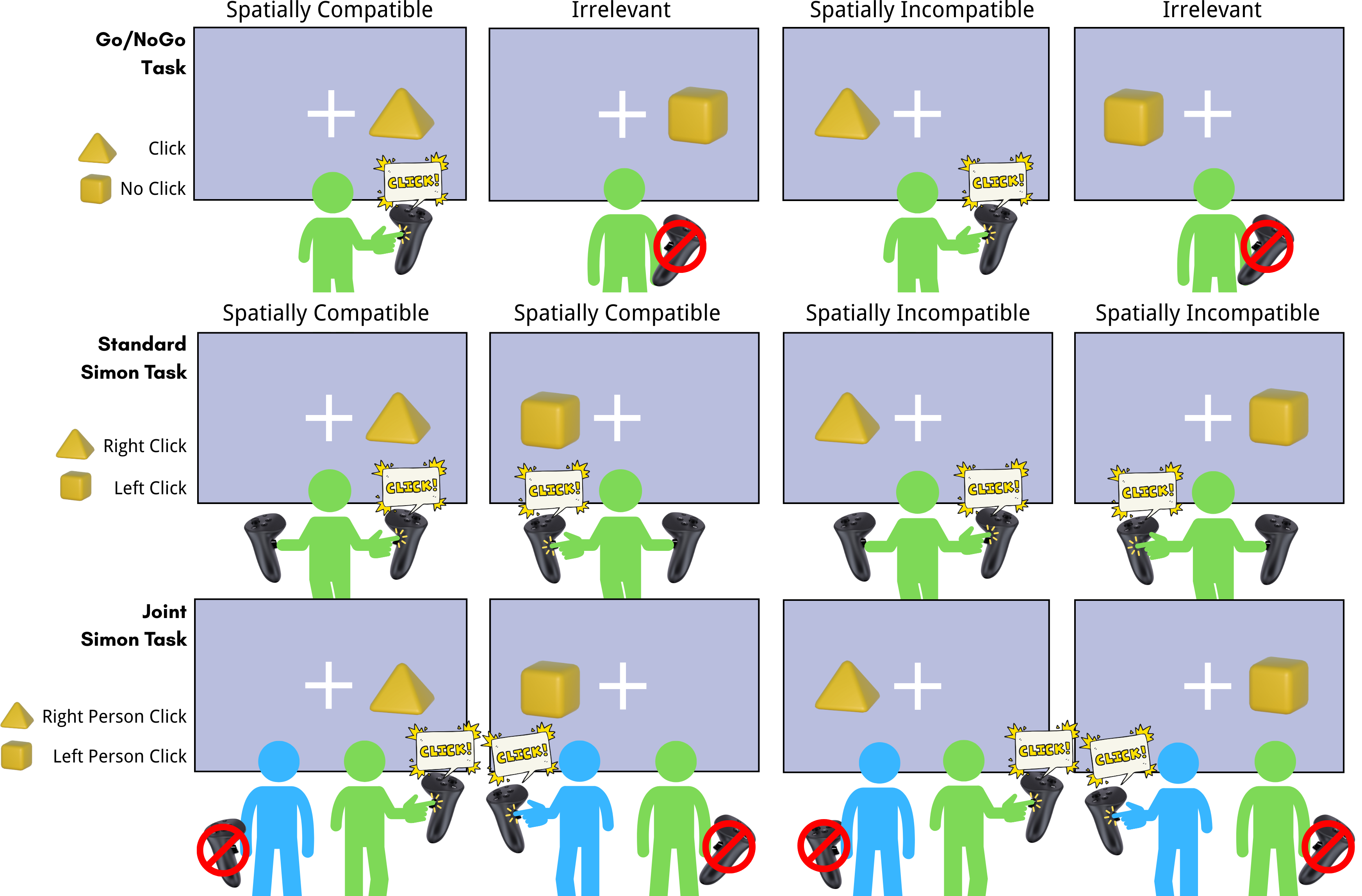}
  \caption{\textbf{Top row}: Go/No-Go Task (single participant). Only one shape (pyramid) is mapped to a response; the alternative shape (cube) requires no action. 
  \textbf{Middle row}: Standard Simon Task (single participant). Both shapes are response-relevant. The participant presses the right-hand grip for triangles and the left-hand grip for cubes. Trials are congruent or incongruent depending on stimulus side. 
  \textbf{Bottom row}: Joint Simon Task. One person responds to triangles, the other to cubes. Compatibility is defined relative to the responding person’s side of space.}
  \label{fig:conditions}
\end{figure*}

The remaining two conditions reintroduced a partner, with and without reciprocal body-swapping, allowing us to test whether the reappearance of spatial compatibility effects results from shared action and altered embodiment rather than extraneous variables. Figure~\ref{fig:conditions} illustrates the stimulus–response mapping for each condition.

\subsubsection{Self-Avatar Embodiment}
Following the approach used by Döllinger et al.\ \cite{dollinger2024virtual}, each experimental block began with a self-avatar embodiment phase designed to help participants familiarize themselves with their avatar. This phase lasted 90 seconds. During this time, participants could lift their feet, swing their arms, turn their heads, and make facial expressions while viewing their avatar from both a first-person perspective and a virtual mirror anchored to the laboratory whiteboard. The mirror, positioned 1.65 meters in front of the standing area, provided a stable third-person view aligned with the stimulus display.

\begin{figure}[h]
    \centering
    \includegraphics[scale=0.17]{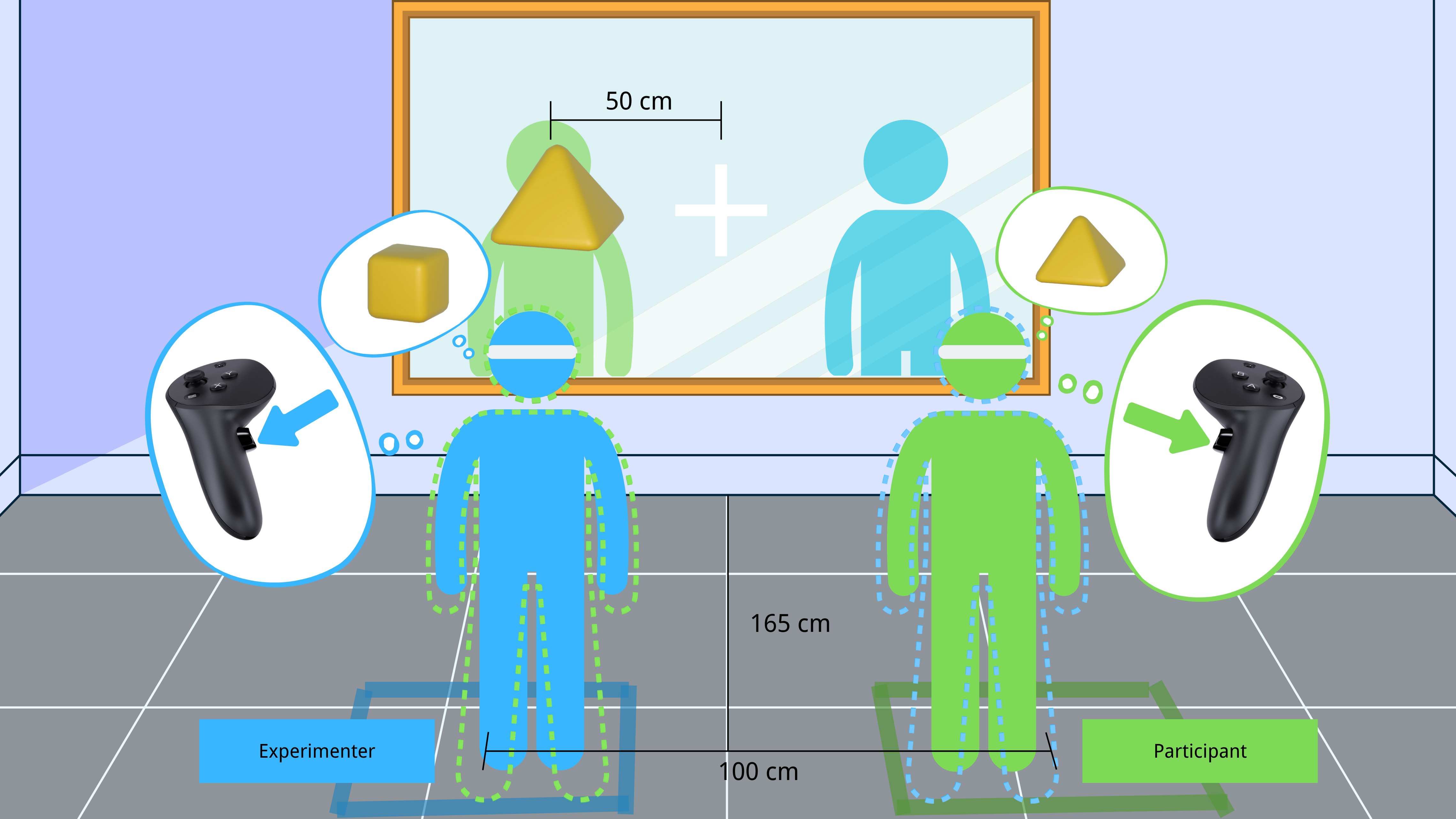}
    \caption{Visualization of the Swapped-Avatar condition. The participant (green) controls the experimenter’s avatar (blue) and responds to the pyramid, which appears in a spatially incompatible position.}
    \label{fig:scene}
    \vspace{-4mm}
\end{figure}

At the end of the self-avatar embodiment, two colored floor markers (0.6 m × 0.6 m) appeared, spaced 1 meter apart and centered along the mirror’s midline: a green square on the participant’s right and a blue one on the left. The participant stood on the green marker throughout all trials, while the partner or experimenter stood on the blue. This arrangement kept both users within the headset’s optimal tracking zone and ensured gaze alignment with the stimulus panel (see Figure~\ref{fig:scene}).

\subsubsection{Standard Simon Block}

Participants performed the Standard Simon Task alone, standing on the green square. They used grip buttons on both left and right controllers to respond to different stimuli (e.g., pressing the left grip for a cube and the right grip for a pyramid). This condition established baseline reaction times (RTs) and accuracy for comparison with subsequent joint tasks. The stimulus–response mapping is illustrated in the middle row of Figure~\ref{fig:conditions}.

\subsubsection{Go/No-Go Block}
Participants performed the Go/No-Go Task alone while standing on the green square. In this condition, they used only the right controller grip button to respond to a target stimulus (e.g., a pyramid) and withheld response for a non-target stimulus (e.g., a cube). This task emphasized impulse control and selective attention. The stimulus–response mapping is shown in the top row of Figure~\ref{fig:conditions}.

\subsubsection{Joint Simon Block (Self-Avatar)}

In this condition, participants used their own avatar and completed a 1.5-minute interaction phase with the experimenter in the virtual environment. This phase allowed both individuals to observe movement synchronization and become familiar with each other’s avatars. They then returned to their assigned positions, with the participant on the green square and the experimenter on the blue.

Each person was assigned a subset of stimuli, such as items appearing on a specific side or in a specific shape, and responded using only the right controller grip button. The task required coordinated responses between partners. The stimulus–response mapping is shown in the bottom row of Figure~\ref{fig:conditions}.

Participants then removed the headset and completed the self-report questionnaires.

\subsubsection{Joint Simon Block (Swapped-Avatar)}

This condition followed the same joint Simon task framework but included a body swap during the interaction phase. After a brief interaction, the experimenter instructed the participant to touch their index finger, triggering the avatar swap. The participant then controlled the avatar previously operated by the experimenter, and vice versa.

Both individuals performed simple movements for 1.5 minutes to acclimate to their swapped avatars before returning to the green and blue squares. The task followed the same structure as in the Self-Avatar condition, with each participant using the right grip button to respond.

Participants then removed the headset and completed the self-report questionnaires.

\subsection{Debriefing}
After completing the questionnaires, the experimenter conducted a short debriefing to discuss the participant’s experience, the mixed reality setup, and avatar control.

\section{Results}

All analyses were conducted in Python 3.11 using pandas 2.2, SciPy 1.13, and statsmodels 0.14. Trials with reaction times (RT) below 150 ms or above 1000 ms (1.8\% of data), as well as incorrect responses (4.1\% on average), were excluded prior to aggregation. Unless otherwise specified, all tests were two-tailed with $\alpha = .05$. Effect sizes are reported as Cohen’s $d$ for $t$-tests and unstandardized $\beta$ for mixed models.

\subsection{Implicit}
We first computed the \textbf{Simon Effect} as the difference in mean reaction time (RT) between spatially incompatible and compatible trials, as shown in Equation~\ref{eq:simon}:
\begin{equation}
\label{eq:simon}
\text{Simon Effect} = \text{MeanRT}_{\text{Incompatible}} - \text{MeanRT}_{\text{Compatible}}
\end{equation}

\subsubsection{Standard Simon Task}
The Standard Simon Task reproduced the classic Simon effect. Mean reaction times (RTs) were $M = 1035.2$\,ms ($SD = 78.6$) for spatially compatible trials and $M = 1087.9$\,ms ($SD = 83.4$) for incompatible trials, $t(31) = 9.33$, $p < .001$, $d = 1.65$, $\Delta = 52.7$\,ms (see Figure~\ref{fig:simon_effect}, left).

\subsubsection{Go/No-Go Task}
\begin{figure}[h]
\centering
\includegraphics[width=\linewidth]{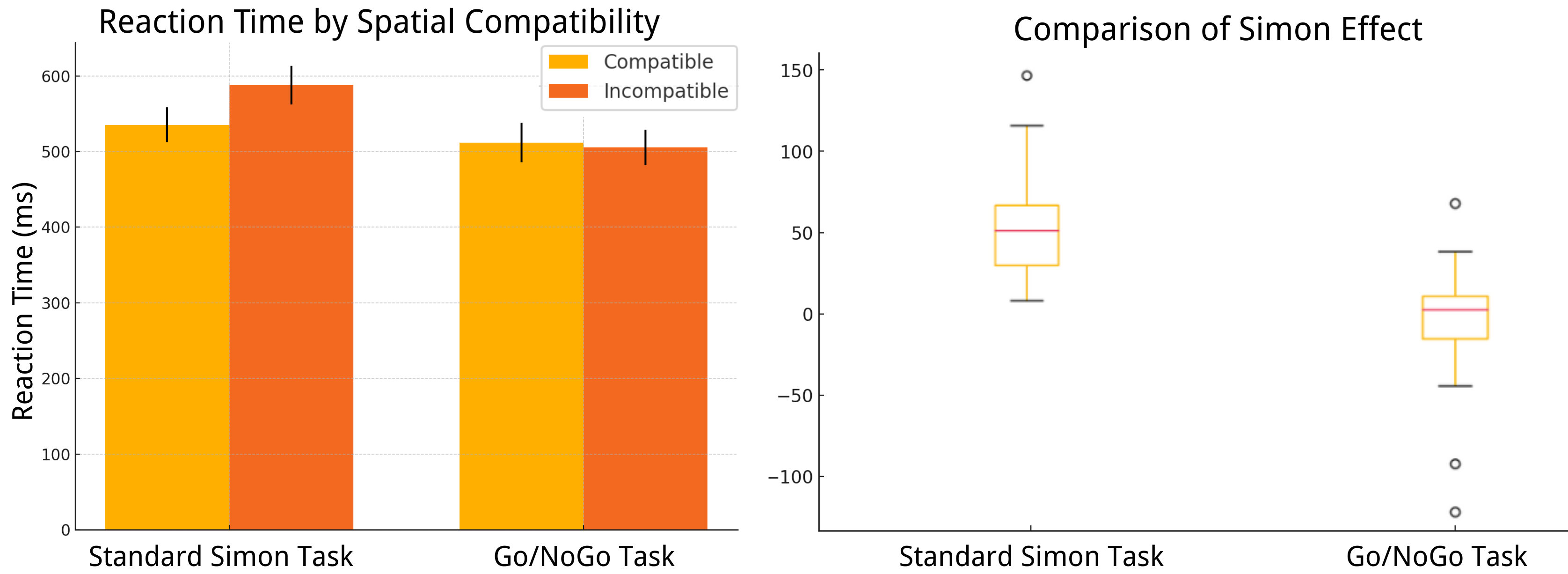}
\caption{Mean RTs for compatible and incompatible trials in the single-participant tasks. \textbf{Left:} Standard Simon Task shows a robust compatibility effect, while Go/No-Go shows minimal RT difference. \textbf{Right:} Boxplot of the Simon Effect (incompatible minus compatible RT) across both tasks.}
\label{fig:simon_effect}
\end{figure}

In contrast, the Go/No-Go Task showed no compatibility effect, $\Delta = -6.2$\,ms, $t(31) = -1.06$, $p = .30$, confirming that using a single response key abolishes the Simon effect (see Figure~\ref{fig:simon_effect}, right).

\subsubsection{Joint Simon Effect}

We next examined how body swapping influenced the \textbf{Joint Simon Effect (JSE)}. For both Joint Simon conditions (Self-Avatar and Swapped-Avatar), we computed the JSE as the mean RT difference between spatially incompatible and compatible trials. We then fit a linear mixed-effects model with \textit{Condition} (Self-Avatar compared to Swapped-Avatar) as a fixed effect and \textit{Participant ID} as a random intercept.

\begin{figure}[h]
\centering
\includegraphics[width=\linewidth]{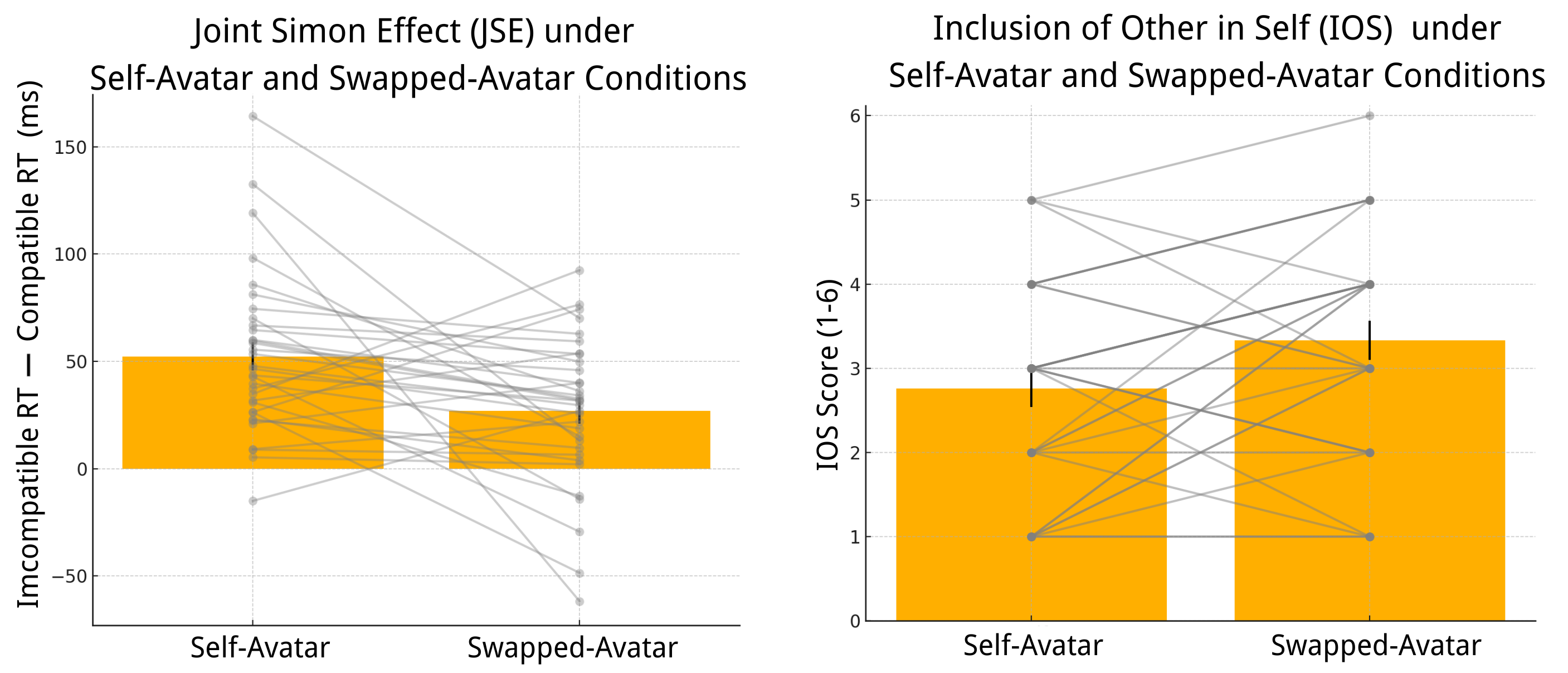}
\caption{\textbf{Left:} Mean Joint Simon Effect (JSE) under the Self-Avatar and Swapped-Avatar conditions. Each gray line represents one participant’s values across the two conditions, with lower JSEs in the Swapped-Avatar condition. Error bars indicate $\pm$1 SE. \textbf{Right:} Mean IOS scores under the same conditions. Each line represents one participant’s ratings, with higher IOS observed after swapping avatars. Error bars indicate $\pm$1 SE.}
\label{fig:jse-ios}
\end{figure}

The model revealed a significant intercept, $b = 52.33$, $\text{SE} = 6.29$, $z = 8.32$, $p < .001$, indicating a robust JSE in the Self-Avatar condition. Critically, the Swapped-Avatar condition showed a significantly reduced JSE, $b = -25.31$, $\text{SE} = 8.66$, $z = -2.92$, $p = .003$, suggesting that embodying a partner’s avatar weakens the spatial interference typically observed when participants respond from their own perspective. Figure~\ref{fig:jse-ios}, left, illustrates this reduction, with each participant’s data connected by a gray line. These results support \textbf{H1}.

\subsection{Experiment Questionnaire Analysis}
We fit a separate linear mixed-effects model for each measure, treating \textit{Condition} (Self-Avatar compared to Swapped-Avatar) as a fixed effect and \textit{Participant ID} as a random intercept.

For perceived interpersonal closeness (IOS), there was a significant main effect of condition, $b = 0.58$, $\text{SE} = 0.24$, $z = 2.41$, $p = .016$, with higher IOS in the Swapped-Avatar condition ($M = 3.34$) than in the Self-Avatar condition ($M = 2.76$). This result indicates that controlling another person’s avatar increased perceived closeness. The right panel of Figure~\ref{fig:jse-ios} displays these IOS differences.

Key estimates for the subscales of Embodiment and Co-Presence are presented in Table~\ref{tab:subjective_measures}. Body ownership was significantly lower in the Swapped-Avatar condition ($\beta = -1.72$, $p < .001$), while self-location also declined slightly ($\beta = -0.35$, $p = .043$). In contrast, co-presence increased ($\beta = +0.34$, $p = .039$), suggesting that participants felt a stronger sense of shared space even while controlling their partner’s avatar. No significant differences were found for agency, attentional allocation, or perceived behavioral interdependence.

\begin{table}[h]
\centering
\small
\caption{Linear Mixed-Effects Model Results for Subjective Experience Measures}
\label{tab:subjective_measures}
\begin{tabular}{lccc}
\toprule
\textbf{Measure} & \makecell{\textbf{Estimate} \\ \textbf{(Swapped - Self)}} & \textbf{p-value} & 95\% CI \\
\midrule
\textbf{Ownership}    & $-1.72^{***}$  & \textbf{$<.001$}   & $[-2.17,\,-1.27]$ \\
Agency               & $+0.02$        & .864      & $[-0.24,\,+0.28]$ \\
Location             & $-0.35^{*}$    & .043      & $[-0.70,\,-0.01]$ \\
\textbf{Co-presence} & $+0.34^{*}$    & \textbf{.039} & $[+0.02,\,+0.67]$ \\
Attentional Allocation & $+0.19$     & .217      & $[-0.11,\,+0.50]$ \\
\textbf{Behavioral Interdependence} & $+0.60$     & \textbf{.003}      & $[+0.20,\,+0.99]$ \\
\bottomrule
\end{tabular}
\end{table}

\subsubsection{Linking IOS and JSE}

Higher IOS scores in the Self-Avatar condition were associated with a stronger Joint Simon Effect. A trial-level mixed-effects model yielded a significant Compatibility × IOS interaction ($\beta = +10.8 \pm 4.5$ ms per IOS point, $z = 2.40$, $p = .019$). A simpler participant-level correlation confirmed the pattern (Spearman $\rho = .34$, $p = .049$). These findings support \textbf{H2a} and are consistent with the proposed link between interpersonal closeness and action co-representation when using self-avatars.

In contrast, this relationship disappeared in the Swapped-Avatar condition. The Compatibility × IOS interaction was not significant ($\beta = -1.2 \pm 4.7$ ms, $p = .80$), and a change-score analysis showed that the IOS increase induced by swapping (ΔIOS) did not account for the corresponding reduction in the JSE ($\beta = -3.0 \pm 6.1$ ms, $p = .63$). These results do not support \textbf{H2b}.

Together, these findings suggest that when body ownership is disrupted, interpersonal closeness no longer modulates spatial interference, underscoring the pivotal role of ownership signals in joint action.

\subsubsection{Linking SoE and JSE}

\begin{figure}[h]
\centering
\includegraphics[width=0.8\linewidth]{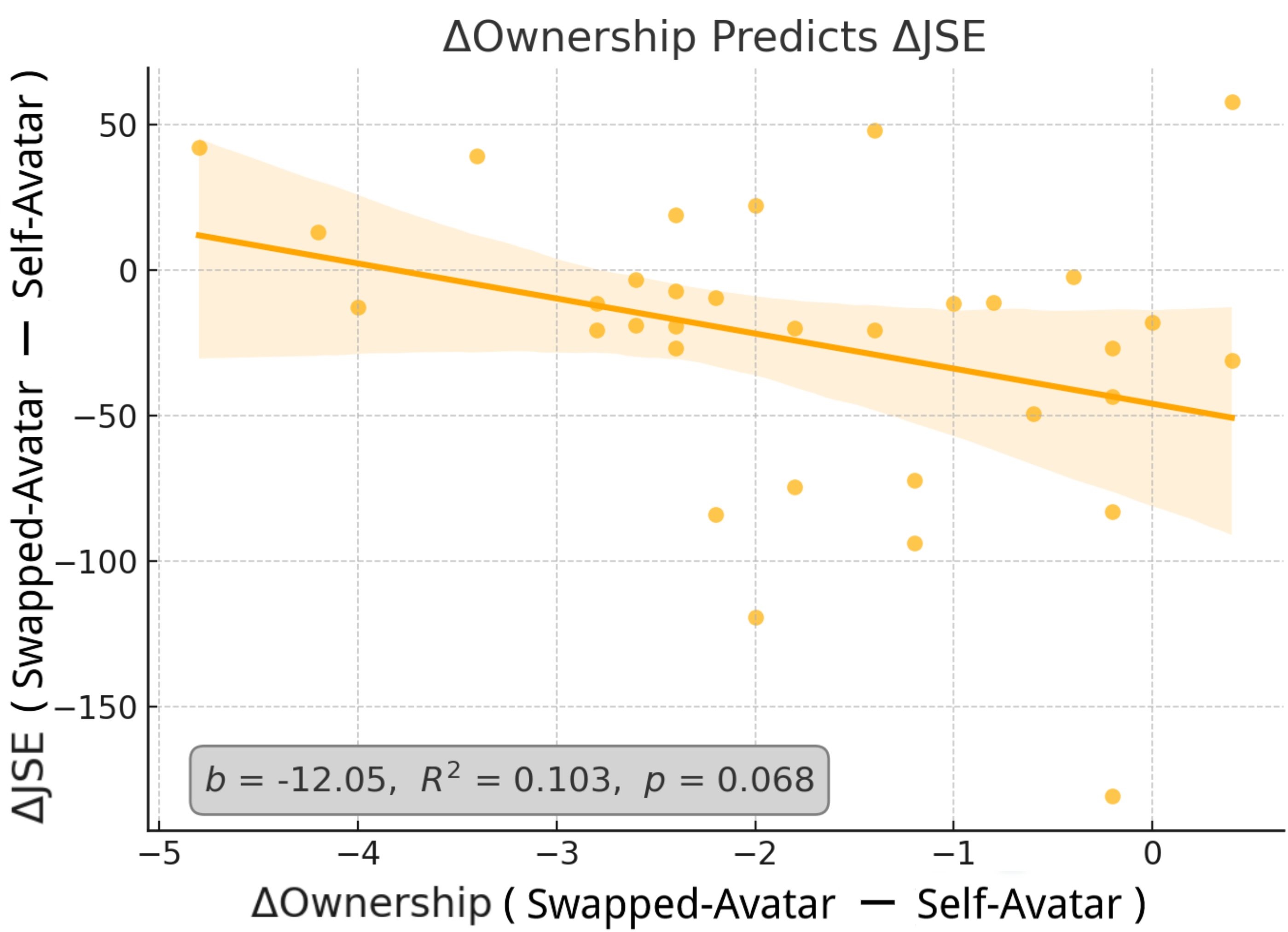}
\caption{Relationship between changes in body ownership and the Joint Simon Effect (JSE). Each dot represents one participant; negative values on both axes indicate that both ownership and the JSE decreased after avatar swapping. The regression line ($\beta = -12$ ms, $p = .068$) indicates that greater ownership loss corresponds to a larger reduction in the JSE. The shaded ribbon represents the 95\% confidence interval.}
\label{fig:ownership}
\end{figure}

To test \textbf{H3a–c}, we examined whether the reduction in the Joint Simon Effect (JSE) following body-swapping was predicted by changes in subcomponents of embodiment or social presence. For each participant, we computed change scores ($\Delta$ = Swapped–Avatar – Self–Avatar) for the JSE and each subjective scale. Ordinary least-squares regressions revealed a marginal negative association between $\Delta$Ownership and $\Delta$JSE ($\beta = -12.1$ ms per scale point, $p = .068$, $R^2 = .10$; see Figure~\ref{fig:ownership}). Participants who experienced greater ownership loss showed a larger reduction in spatial compatibility, partially supporting \textbf{H3a}.

Parallel analyses found no significant predictive effects for changes in agency, self-location, interpersonal closeness (IOS), co-presence, attentional allocation, or behavioral interdependence (all $|\beta| < 6$ ms, $p > .25$). These null results do not support \textbf{H3b} or \textbf{H3c}, suggesting that behavioral decoupling is specifically linked to diminished body ownership rather than general shifts in social presence or other embodiment components.

A within-subject mediation analysis, placing $\Delta$Ownership between Condition and $\Delta$JSE, yielded a marginal indirect effect ($-6.3$ ms, 95\% CI [$-14.8$, $0.2$], $p = .062$), suggesting that ownership loss may partially account for the observed attenuation.

\subsection{Trait Questionnaire Analysis}
We next examined whether stable individual differences moderated the subjective and behavioral outcomes of body-swapping (\textbf{H4}). For each participant, we regressed the change in interpersonal closeness ($\Delta$IOS = IOS$_\text{Swapped-Avatar}$ – IOS$_\text{Self-Avatar}$) on each trait score.

\begin{figure}[h]
\centering
\includegraphics[width=\linewidth]{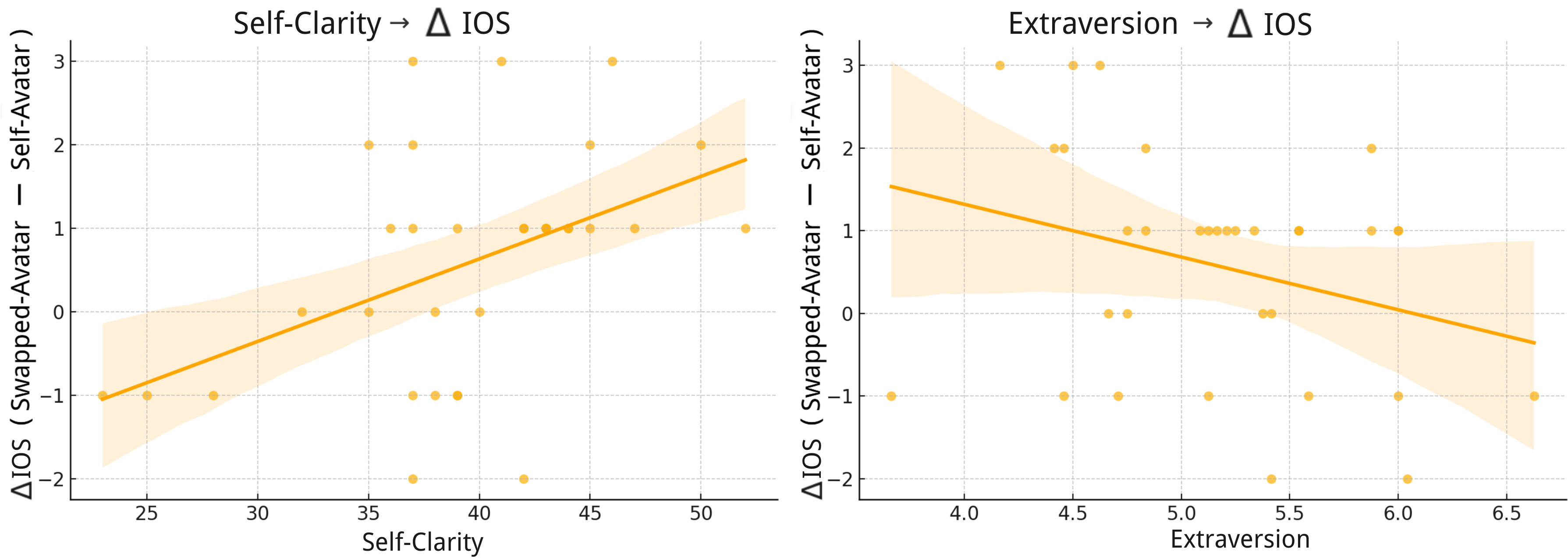}
\caption{\textbf{Trait predictors of $\Delta$IOS}. \textbf{Left:} Higher Self-Concept Clarity (SCC) is associated with a larger IOS increase after avatar swapping ($\beta = 0.10$, $p = .008$). \textbf{Right:} Extraversion shows a non-significant negative trend ($\beta = -0.64$, $p = .09$). Shaded ribbons indicate 95\% confidence intervals.}
\label{fig:trait-ios}
\end{figure}

A simple least-squares regression revealed a significant positive association with Self-Concept Clarity (SCC), $\beta = +0.10 \pm 0.03$ IOS units per SCC point, $t(31) = 2.98$, $p = .008$, $R^2 = .22$. Participants with a clearer, more coherent self-concept showed a greater increase in perceived closeness under the Swapped-Avatar condition (Figure~\ref{fig:trait-ios}, left).

Among the Big Five traits, Extraversion showed a marginal negative trend, $\beta = -0.64 \pm 0.37$, $t(31) = -1.75$, $p = .090$, $R^2 = .08$. More extraverted individuals tended to report slightly smaller IOS increases (Figure~\ref{fig:trait-ios}, right). Agreeableness, Conscientiousness, Emotional Stability, and Openness were not significantly related to $\Delta$IOS ($|\rho| < .17$, $p > .35$). No trait score significantly moderated the reduction in the Joint Simon Effect ($p > .30$ for all Condition × Trait interactions).

\subsection{Summary}
Reciprocal body swapping in mixed reality reduced the Joint Simon Effect (JSE) while increasing interpersonal closeness, as measured by the Inclusion of Other in the Self (IOS) scale. This IOS increase was more pronounced in participants with higher self-concept clarity. In contrast, the reduction in JSE was associated with the extent of body ownership loss. No other subjective experiences or personality traits showed significant associations with these outcomes. These findings suggest that virtual body swapping can reshape how individuals co-represent actions during collaborative tasks.

\section{Discussion}

\subsection{Joint Simon Effect Attenuation through Body Swapping}

The present study replicated the classic Joint Simon Effect (JSE) in mixed reality (MR): participants responded significantly more slowly on spatially incompatible trials than on compatible ones. However, this effect was markedly reduced when participants controlled their partner’s avatar. This attenuation suggests that reciprocal body swapping alters the spatial coding mechanisms underlying the JSE. Specifically, when participants controlled another body while seeing their own avatar controlled externally, the usual self–other action distinction that drives spatial interference was weakened. This disruption offers insight into how altered embodiment affects joint action representation.

\textbf{Predictive processing models} of embodiment propose that body ownership is sustained only when incoming multisensory signals match prior expectations about the body \cite{kilteni2012sense}. The mismatch introduced by body swapping weakens this prior, leading participants to feel less certain that either avatar belongs to them.

Our findings further support the idea that, during body swapping, components of embodiment are functionally distributed across avatars. Visual appearance, typically contributing to ownership, is assigned to the partner’s avatar, while motor control, associated with agency, remains with the user. This division may promote deeper self–other integration, encouraging participants to perceive both avatars as a unified embodied system.

\textbf{Referential coding theory} of the JSE posits that spatial interference arises when competing actions are categorized as self-generated versus other-generated \cite{dolk2014joint}. Once ownership is diminished, the partner’s response no longer functions as an external cue, and spatial incompatibility loses diagnostic value, thereby contracting the JSE.

Debriefing comments echoed this interpretation. One participant remarked, “\emph{My avatar feels like part of me, so when someone else takes it over I feel resistance; using a body that isn’t mine feels uncomfortable}.” Another said it was “\emph{strange to watch someone who looks exactly like me doing things I would never do}.” Both comments reflect the expected multisensory prediction error: when participants strongly identified with their avatar, seeing it controlled by someone else or behaving unexpectedly weakened the sense of ownership. These impressions closely align with the observed negative association between $\Delta$Ownership and $\Delta$JSE.

\subsection{The Dissociation of JSE and IOS}
Different from previous studies that typically found a strong association between the Joint Simon Effect (JSE) and interpersonal closeness \cite{shafaei2020effect,ford2015exploring}, our results revealed a dissociation. Although interpersonal closeness (IOS) increased after the body swap, only the loss of ownership significantly predicted the reduction in the JSE. This seemingly paradoxical outcome, in which social merging becomes stronger while spatial conflict diminishes, reflects two distinct levels of self–other processing. At a higher level, interpersonal affiliation intensifies because partners quite literally take each other’s perspective. At a lower level, however, spatial coding breaks down because the bodily reference point that usually distinguishes one’s own actions from those of the other has been disrupted.

\subsection{Individual-Difference Modulation}

Participants with higher scores on the Self-Concept Clarity (SCC) scale exhibited greater increases in interpersonal closeness (IOS) following the Swapped-Avatar condition. A coherent self-concept may provide a stable psychological foundation that supports perspective-taking without disorientation. This finding aligns with self-expansion theory, which holds that individuals with well-defined identities are more receptive to incorporating positive attributes from others.

A marginal negative association was observed for Extraversion, with more extraverted participants reporting smaller IOS increases. This pattern may reflect a ceiling effect, where individuals who already feel socially connected show limited additional gains. Other traits, including Agreeableness, Conscientiousness, Emotional Stability, and Openness to Experience, were not significantly related to IOS change and did not moderate the effect of swapping on the Joint Simon Effect (JSE).

These results suggest a two-part mechanism. Gains in interpersonal closeness depend on the stability of self-concept, whereas the translation of that closeness into coordinated action depends on the integrity of body ownership. For designers of virtual systems, this implies that perspective-taking interventions are most effective when users first establish a stable sense of self. Once that foundation is set, embodiment parameters can be adjusted to balance empathy and sensorimotor coordination. Theoretically, these findings highlight that stable traits and transient embodied states contribute to self–other integration in distinct but complementary ways.

\subsection{Practical Implications for Multi-User XR}
Reciprocal body swapping enhanced interpersonal closeness while reducing action-level coordination. This trade-off offers useful guidance for designing multi-user extended reality (XR) systems.

\paragraph{Collaboration and co-located teamwork.}
In creative or conflict-resolution contexts, design goals often prioritize empathy over precise temporal synchrony. Our findings show that even a short full-body swap increases interpersonal inclusion, particularly among users with a stable self-concept. Designers can introduce brief reciprocal swaps as ice-breakers, allowing users to inhabit each other’s avatars for 2–3 minutes before returning to their original bodies. Since body ownership loss weakens the Joint Simon Effect, tasks requiring spatial coordination (such as object assembly) should resume in the original embodiment to maintain both social connection and performance accuracy.

\paragraph{Therapeutic and well-being applications.}
For clients with a rigid or negative body image, such as those experiencing social anxiety or body dysmorphia, body swapping may help soften bodily boundaries. Since IOS gains were more pronounced in individuals with higher self-concept clarity, clinicians may begin with narrative or cognitive-behavioral techniques to help establish a coherent sense of self before introducing full avatar swaps. For individuals who already exhibit weak bodily boundaries, gentler approaches such as synchronous touch without a perspective shift may offer therapeutic benefits while maintaining the sense of ownership.

\paragraph{Public speaking and performance coaching.}
A novice speaker could embody a confident avatar, while a coach temporarily takes the learner’s perspective to demonstrate gestures, posture, or breathing from first-person view. The resulting IOS increase may build trust and reduce anxiety, making feedback more effective.

\paragraph{Adaptive embodiment cues.}
The connection between body ownership and the JSE suggests new strategies for adaptive embodiment. XR systems could track visuo-proprioceptive alignment and dynamically adjust avatar opacity, outlines, or haptics to reinforce ownership when precise coordination is needed. In socially focused settings such as team-building, reducing ownership cues may promote perspective-taking and empathy.

\medskip
In summary, effective multi-user XR design should increase body swapping to promote empathy, reduce it when control is essential, and dynamically adjust embodiment cues to balance social bonding with coordination demands.

\subsection{Limitations and Future Directions}
Several limitations merit consideration. First, although our sample size (N = 33) was adequate to detect main effects, it limited our ability to capture finer personality-related interactions. Second, while avatar embodiment sessions were brief, future studies could explore whether longer or repeated swaps lead to stronger or qualitatively different outcomes. Third, the study focused on dyadic interactions, typically between a participant and an experimenter. Research involving larger groups or human–agent contexts may reveal more complex self–other dynamics. Finally, we relied on reaction-time and self-report data; adding neurophysiological measures (e.g., EEG, EMG, fNIRS) could yield deeper insights into the mechanisms of ownership, agency, and co-representation.

\section{Conclusion}

This study shows that a fully reciprocal body swap in co-located mixed reality reshapes the self–other boundary along two dimensions. At the sensorimotor level, it attenuates the Joint Simon Effect (JSE), with the degree of reduction closely tied to body ownership loss. At the social level, it increases interpersonal closeness, especially among individuals with a coherent self-concept.

By dissociating these facets of self–other integration, the findings refine theories of referential coding and embodiment. They demonstrate that increased social closeness does not necessarily imply stronger action co-representation when the bodily anchor distinguishing self from other is disrupted.

Methodologically, this is the first implementation of the Joint Simon Task in mixed reality with real-time reciprocal avatar swapping. The study also confirms that millisecond-level reaction time data can be reliably collected under dynamic embodiment.

Practically, the results offer a design principle for multi-user XR: brief or partial body swaps can promote empathy and bonding, while consistent ownership cues should be maintained when precise coordination is required.

Future work may extend this space by examining how avatar similarity, group dynamics, or physiological synchrony shape shared and distributed embodiment. Such efforts can help determine when altered embodiment enhances rather than impairs collaborative performance.

\acknowledgments{
The authors thank Jackey Chai, Rosie Connolly, and Charlotte Dubosc for their generous assistance with video shooting and pilot studies. We are also grateful to all participants for their time and engagement. 
This work was supported by the Science Foundation Ireland Centre for Research Training in Digitally-Enhanced Reality (d-real) under Grant No. 18/CRT/6224.}

\bibliographystyle{abbrv-doi}

\bibliography{Reference}
\end{document}